\newif\ifdraft

\newif\ifcreatebib
\createbibfalse

\ifdraft
\documentclass[preprint,showpacs,preprintnumbers,amsmath,amssymb,floatfix,aps,pre]{
revtex4-1}
\else
\documentclass[twocolumn,showpacs,preprintnumbers,amsmath,amssymb,aps,pre]{revtex4-1}
\fi

\usepackage[utf8x]{inputenc}
\usepackage{palatino}

\usepackage{graphicx}
\usepackage[caption=false]{subfig} 
\usepackage{dsfont}
\usepackage{bm}

\newcommand{\ab}{{\alpha\beta}}

\long\def\comment#1{}
\begin{document}
\preprint{APS/preprint}

\title{Multi-Scale Codes in the Nervous System:\\
The Problem of Noise Correlations\\
and the Ambiguity of Periodic Scales}
\author{Alexander Mathis}
\email{mathis@biologie.uni-muenchen.de}
\author{ Andreas V. M. Herz} 
\author{Martin B. Stemmler}
\affiliation{Bernstein Center for Computational Neuroscience,
Ludwig-Maximilians-Universit\"at M\"unchen}
\date{\today}

\begin{abstract}
Encoding information about continuous variables using noisy computational units is a
challenge; nonetheless, asymptotic theory shows that combining multiple periodic
scales for coding can be highly precise despite the
corrupting influence of noise~\cite{Mathis2012b}. Indeed, cortex seems to use such
stochastic multi-scale periodic `grid codes' to represent position accurately. We
show here how these codes can be
read out without taking the asymptotic limit; even on short time scales, the
precision of neuronal grid codes scales exponentially in the number $N$ of neurons.
Does this finding also hold for neurons that are not statistically independent? To
assess the extent to which biological grid codes are subject to statistical
dependencies, 
we analyze the noise correlations between pairs of grid code neurons in behaving
rodents. We find that if the grids of
the two neurons align and have the same length scale, the noise correlations between
the neurons can reach $0.8$. For increasing mismatches between the grids of the two
neurons, the noise correlations fall rapidly. Incorporating such correlations into a
population coding model reveals that the correlations lessen the resolution, but the
exponential scaling of resolution with $N$ is unaffected.
\end{abstract}

\maketitle

\section{Introduction}

 Multi-scale basis functions, such as simple Fourier transforms or wavelets, have a
long history, dating back to the 19th century. 
 They are widely used for data compression, processing and
analysis~\cite{Haar1909,Chui1992,Unser1996,Mallat2009}. For instance, 
 state-of-the-art image compression algorithms convolve images with a discrete cosine
or Haar transform at different length scales~\cite{Acharaya2005} . 
  Wavelets at multiple scales or  ``steerable
pyramids"~\cite{Simoncelli_etal92,Riesenhuber_Poggio99} are used both in machine and
biological vision.
Indeed, the receptive fields  observed in the early visual
system~\cite{Hubel1968,Devalois1990} resemble wavelets; moreover, they emerge
naturally 
in optimally sparse codes for the visual and auditory
systems of mammals~\cite{Olshausen1996,Lewicki2002}. 

In the last decade, neuroscientists working in the medial entorhinal cortex (mEC),
pre- and parasubiculum have discovered periodic neuronal  tuning
curves~\cite{Hafting2005,Boccara2010}
for stimuli that are not intrinsically periodic.  Neurons with such tuning curves
fire spikes  at regularly spaced locations within an environment. The resulting map
of spatial firing resembles a  hexagonal grid,  inspiring the researchers to call
these neurons `grid cells'. The grids within the same cortical area have a finite
number of different length scales and  the ratio of one length scale to
the next shorter seems to be constant ~\cite{Barry2007,Stensola2012},
in accordance with optimal coding theory~\cite{Mathis2012a,Mathis2012b,Wei2013}.

What role do  periodic tuning curves and multiple scales play in coding?
Recently,  we computed the Fisher information for population codes with such
properties  and showed 
that  their precision can scale exponentially in the number of
neurons~\cite{Mathis2012a,Mathis2012b}, as long as the neurons fired independently.
In contrast, 
the precision of population codes with unimodal or sigmoidal tuning curves
scale linearly in the number of
neurons~\cite{Seung1993,Zhang1999,Wilke2001,Bethge2002,Brown2006,Mcdonnell2008,
Nikitin2009}.  In this paper, we address two major concerns
that might affect the feasibility of multi-scale codes:
\begin{itemize}
\item
 Tuning curves with multiple peaks compound the ambiguity already inherent in a
stochastic representation; this ambiguity can lead 
 to catastrophic decoding errors. As the Fisher information is a local, asymptotic
measure of coding accuracy,  we had provided
 quantitative bounds   for the probability of catastrophic errors. Here we show how
the probability of a stimulus $x$ given the population response 
can be calculated analytically. This allows us to estimate the true coding error,
even  for high-dimensional stimuli and population responses that are sampled
only for short time periods.
\item 
Neurons from the same area in cortex  at times display correlated fluctuations that
are unrelated to the stimulus being encoded.
Such  noise correlations  can be detrimental to the encoding accuracy
of population codes~\cite{Zohary1994,Schneidman2006,Ecker2010,Cohen2011}.
Much of the experimental~\cite{Zohary1994,Lee1998,Smith2008,Ecker2010,
Cohen2011,Miura2012} and theoretical
work~\cite{Abbott1999,Shamir2001,Wilke2001,Shamir2006,Ecker2011} has 
focused on sensory and motor areas of the brain. In contrast, brain regions, such
as the hippocampus and
medial entorhinal cortex (mEC) are less
well studied in this context. As these areas form a central hub
of computation, receiving and sending information from many other brain areas,
correlated fluctuations 
might be a byproduct of the neuronal network's processing.  
Here we quantify the noise correlations of grid cells in the entorhinal cortex (EC)
of
behaving rodents and study their effect on the coding accuracy.
\end{itemize}

\begin{figure}[!ht]
\begin{center}
\includegraphics[angle=0,width=0.8\columnwidth]{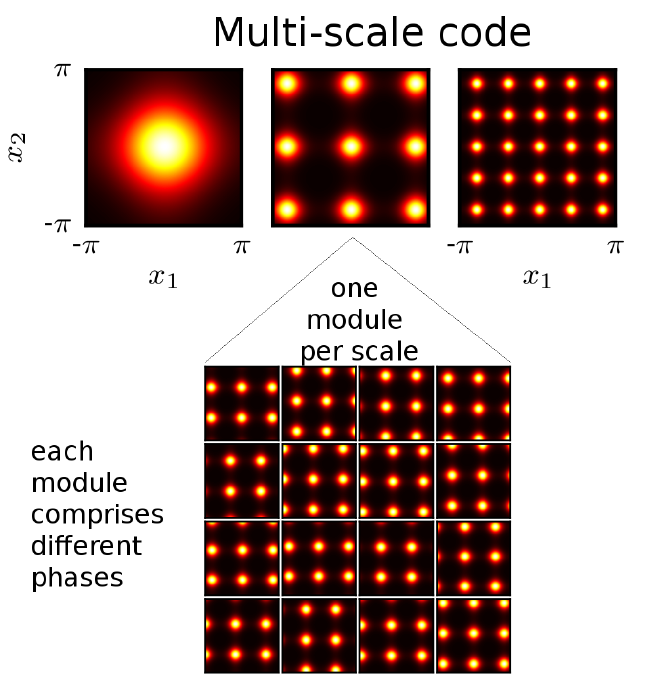}
\end{center}\caption{Illustration of a multi-scale code. The population consists of
neurons that encode the stimulus with tuning curves $\Omega_i(x)$ that vary in
scale and whose amplitude is shown in color. Each scale comprises a group of cells
with periodic tuning curves on an 
identical lattice
structure but different shifts (phases) -- an arrangement we call a module. Such a
population of nested modules successively refines the representation of the stimulus.
For a single module, the mapping from
stimulus to periodic population response is not injective; only the response over the
whole ensemble of modules provides a unique representation of the
stimulus within $[-\pi,\pi)^2$.}
\label{fig:Introfigure}
\end{figure}

The paper is organized as follows. First, we compute the maximum likelihood estimate
of the stimulus 
from the neuronal response, and compare the resulting error to the prediction from
the Fisher information. 
Second, we analyze the effect of  noise correlations on the Fisher information in
multi-scale population codes and compare
these results to numerical estimates of the mean square error (MSE).  
Third, we analyze the noise correlations between grid cells from real data (provided
by the Moser lab at the Norwegian
University of Science and Technology~\cite{Hafting2008}) to corroborate
the correlated noise model we used. We find that although noise
correlations reduce the overall accuracy, multi-scale codes, as found in mEC, are
still vastly superior to single-scale codes.

\section{The Precision of Multi-scale Codes---Long-Term Asymptotics versus Short-Term
Estimates}

Take a population of $N$ neurons.  In response to a stimulus $\bm x \in I \subset
\mathds{R}^D$, the activity  of the $i$-th neuron is 
\begin{equation}
\Upsilon_i(\bm x) = \Omega_i(\bm x) + \eta_i(\bm x). \label{model}
\end{equation}
Here $\Omega_i(\bm x)$ is the average response of neuron $i$, also known as the
neuron's tuning curve,  while $\eta_i$ represents
the trial-to-trial variability.  The variability $\eta_i(\bm x)$ has zero mean by
definihtion, but may be correlated
across neurons--- a case that we treat in the next section.

In previous work~\cite{Mathis2012b,Mathis2012a}, we argued that
a population code should consist of periodic, multi-scale tuning curves $
\Omega_i(\bm x)$, as exemplified
by the \textit{von Mises functions} in one dimension:
\begin{equation}
\Omega_i(x) = f_{max} \tau  \cdot \exp \left( \frac{\cos(2 \pi/\lambda_i (x
- \varphi_i))-1}{\sigma^2}\right),\label{tuningcurves}
\end{equation}
as illustrated in Fig.~\ref{fig:Introfigure}.
The tuning curve $\Omega_i(\bm x)$ of each cell $i$ is characterized by a scale
$\lambda_i$, a preferred phase $\varphi_i$ and tuning width $\sigma$. The term
$f_{max}$ denotes the peak firing rate (number of action potentials per unit time),
which is assumed to be the identical for all neurons, and $\tau$ describes the
characteristic time interval over which spikes are counted.
These tuning curves would ideally be organized into different, nested modules such
that all neurons within one module share the period $\lambda_i$ but exhibit different
preferred phases $\varphi_i$. Such multi-scale population codes achieve exponentially
higher precision in
representing $x$ than unimodal codes,
provided that the $\lambda_i$'s are arranged into a discrete, geometric
progression~\cite{Mathis2012a}. 
Recent experimental results on grid codes in entorhinal cortex bear out this
theoretical prediction~\cite{Barry2007,Stensola2012}.

Given a stimulus $x$, the response has a probability distribution $P({\bm \Upsilon} |
\bm x)$, where $ \bm \Upsilon= (\Upsilon_1, \dots, \Upsilon_N)$ denotes the
population's response.
The task for an ideal observer is to estimate $x$  from $\bm \Upsilon$ as $\hat{\bm
x}$, 
for instance by choosing the most likely stimulus, or the one that minimizes
${\left\langle (\bm x-\hat{\bm x})^2\right\rangle}_{p(\bm x)}$. 
Asymptotically, as ${f_{max} \tau} \to \infty$, a statistically efficient estimator
$\hat{\bm x}$ will have a probability distribution that approaches
\begin{equation}
p( \hat{\bm x}-\bm x) \sim \exp\left[ -  \frac{1}{2}
{\left(  \hat{ \bm x}-\bm x\right)}^T {\mathbf J} \left( \hat{ \bm x}- \bm x\right)
\right],
\label{eq:probxhat}
\end{equation}
at least as long as $\hat{\bm x}$ is close to the true $\bm x$. Here $\mathbf{J} $ is
the Fisher information matrix at position $x$ with entries
\begin{equation}
\bm J_{\ab}(\bm x) = \left\langle  \left(\frac{\partial  \ln P(\bm \Upsilon|\bm
x)}{\partial x_\alpha}
\right)
\left(\frac{\partial \ln P(\bm \Upsilon|\bm x)}{\partial x_\beta}\right)
\right\rangle_{P(\bm \Upsilon|\bm x)},
\end{equation} 
where $\alpha$, $\beta \in \{1,\cdots, D\}$. Indeed, the Cram\'er-Rao bound strictly
limits the error of any unbiased estimate 
of $x$ through the Fisher information
\begin{equation}
  \left\langle (\bm x-\hat{\bm x})^2\right\rangle \geq \bm J(\bm x)^{-1}.
 \label{eq:cramer-rao}
\end{equation}
The key question is: how close will an efficient estimator come to the Cram\'er-Rao
bound?  In a multi-scale, periodic grid code, $p\left( \bm x-  \hat{\bm x}\right)$
will deviate
from the Laplace approximation inherent in Eq.~\eqref{eq:probxhat}; the periodicity
causes the distribution $p\left(\bm x- \bm  \hat{\bm x}\right)$  to have multiple
peaks. We now show how one 
can avoid the assumption of the asymptotic limit entirely, by extending a result of 
Yaeli and Meir~\cite{Yaeli2010} on Gaussian tuning curves.

\begin{figure}[!ht]
\begin{center}
\ifdraft
\includegraphics[angle=0,width=0.6\columnwidth]{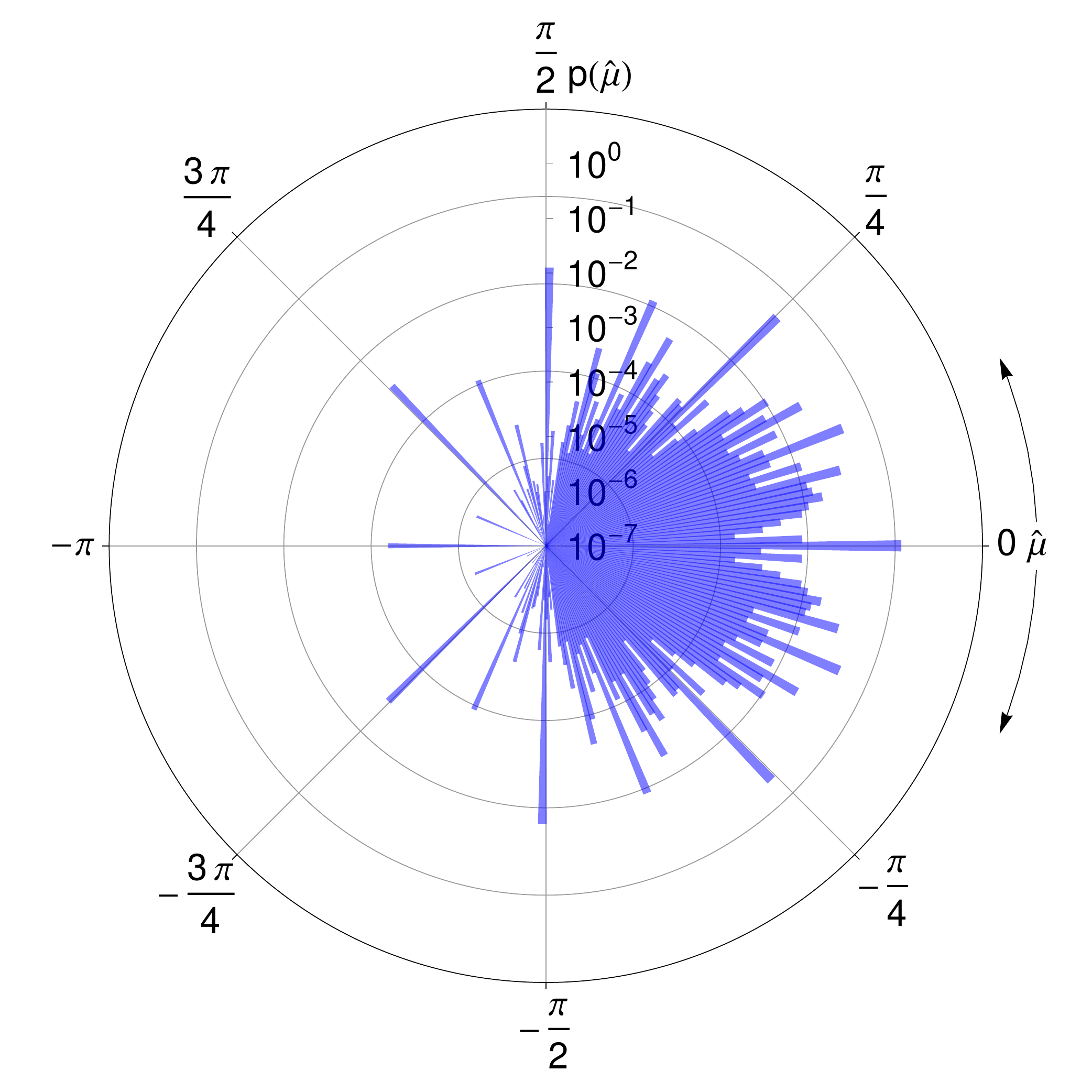}
\includegraphics[angle=0,width=0.6\columnwidth]{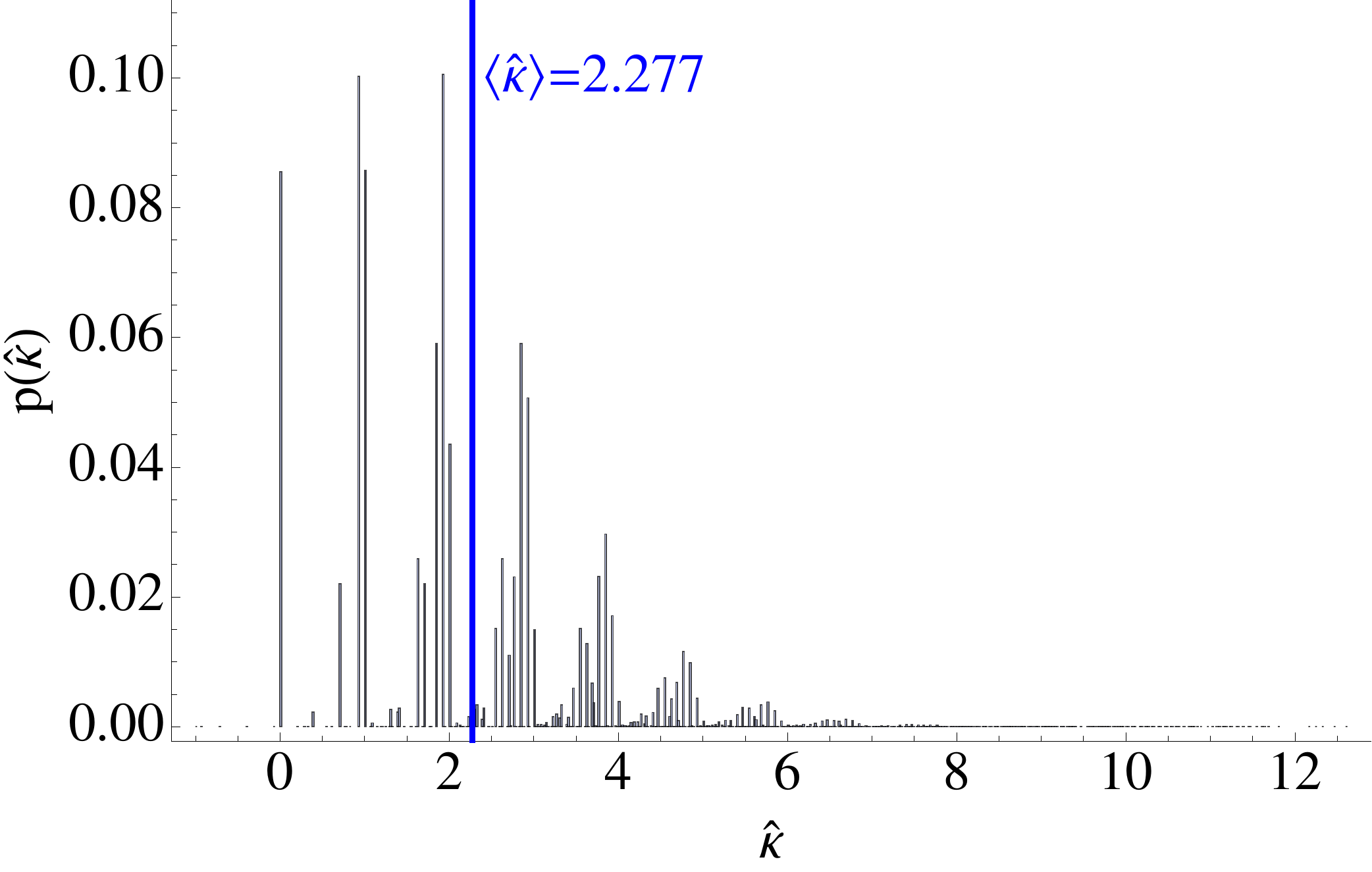}
\else
\includegraphics[angle=0,width=\columnwidth]{Fig2a.pdf}
\includegraphics[angle=0,width=\columnwidth]{Fig2b.pdf}
\fi
\end{center}\caption{Distribution of estimates $\hat{\mu}$ and
$\hat{\kappa}$: As the population response $\bm \Upsilon$ is stochastic, the
posterior distribution $P(x|\bm \Upsilon)$ depends on the random
variables $\hat{\mu}$ and $\hat{\kappa}$, which govern the {\it von Mises}
distribution in Eq.~\eqref{eq:posterior_von_Mises}. The probability distributions of
these two 
random variables are shown above for a  single-scale module of neurons with
$f_{\text{max}} \tau=1$, $M=16$, $\lambda_0= 2 \pi$ and $\sigma^2=1/7$. The quantity
$\hat{\mu}$ is the maximum a posteriori (MAP) estimate of $x$. 
The asymptotic Fisher information $J$ is proportional to the expected value of
$\hat{\kappa}$. For a given realization of $\bm \Upsilon$,
the expected error can be greater or less than predicted by the asymptotic
Cram{\'e}r-Rao bound, depending on whether $\hat{\kappa}$ is less or greater than
$\left\langle 
\hat{\kappa} \right\rangle$. If $\hat{\kappa} \gg 1$, then the expected error is
$1/\hat{\kappa}$, and the Cram\'er-Rao bound for the average error is 
a consequence of Jensen's inequality,  $\left\langle 1/\hat{\kappa}\right\rangle \geq
1/\left\langle \hat\kappa\right\rangle = 1/J$.  }
\label{fig:pmu_and_pk}
\end{figure}

For simplicity, let us start with one module of $M$ neurons, each with a tuning curve
$\Omega_i(x)$ given by Eq.~\eqref{tuningcurves} 
on 
the one-dimensional interval $I = [-\pi,\pi)$. These tuning curves have a  uniform
period $\lambda_0=2 \pi $ and tuning width 
$\sigma^2$, but the parameter $\varphi_i$ is distributed across different neurons, so
that
the tuning curves cover the interval uniformly. 
The prior probability of $p(x)$ is assumed to be uniform, and each neuron's
response
$\Upsilon_i$ obeys a discrete Poisson distribution.
 If the neurons are statistically independent,
then by Bayes' rule
\begin{align}
p\left( x | {\bm \Upsilon}  \right)   \sim    & \prod_{j=1}^M   \text{Poisson} \left(
\Upsilon_j, \Omega_j(x) \right) \notag  \\
=  & \frac{  \prod_{j=1}^M  \exp( \Upsilon_j \ln \left(\Omega_j(x)\right)} 
{ \prod_{j=1}^M \Upsilon_j !} \exp \!\!  \left(\!\! - \sum_{j=1}^M \Omega_j(x) \!\!
\right) \notag \\
\intertext{As observed by several authors~\cite{Dayan2001,Yaeli2010}, if the tuning
curves uniformly  cover the interval, 
$\sum_{j=1}^M  \Omega_j(x)  \approx \text{constant}$, even for relatively small $M$.
With this one approximation,}
P\left( x | {\bm \Upsilon }\right) = & C \cdot  \exp \left( \kappa \sum_{j=1}^M
\Upsilon_j \cos( x-\varphi_j ) \right),
\label{eq:posterior}
\intertext{where $C$ is a normalization constant, and $\kappa = \sigma^{-2}$.
We can express this  posterior probability  as a von Mises function with mean
$\hat{\mu}$ and concentration 
$\hat{\kappa}$}
P(x| \bm \Upsilon) = & C \exp\left\{\hat{\kappa} \cos \left[  (x - \hat{\mu})\right]
\right\}.
\label{eq:posterior_von_Mises}
\end{align}
Within the interval $[-\pi,\pi)$, $x=\hat{\mu}$ corresponds to the peak of
Eq.~\eqref{eq:posterior}, so that $\hat{\mu}$ represents the {\em most likely} 
stimulus, given the neurons' response. The concentration $\hat{\kappa}$ expresses
the certainty about $\hat{\mu}$. We obtain
\begin{align*}
\hat{\mu} =  &
 \text{arg} \left( \sum_{j=1}^M \Upsilon_j \exp( i   \, \varphi_j  )\right)\\
 \intertext{ and } 
\hat{\kappa} = &  \kappa \sum_{j=1}^M \Upsilon_j  \cos \left[ \left( \hat{\mu} -
\varphi_j
\right) \right].
\end{align*}
So the expected phase $\hat{\mu} \in [-\pi,\pi)$ is the response-weighted sum of the
neurons' preferred phases, also known as the population
vector~\cite{Georgopoulos1986,Seung1993}.
Both $\hat{\kappa}$ and $\hat{\mu}$ are random variables, as the population's
response  vector $\bm \Upsilon$ is stochastic (see Fig.~\ref{fig:pmu_and_pk}).
The expected value of $\hat{\kappa}$ is 
$$
\left\langle \hat{\kappa}\right\rangle =   \kappa   \int_{-\pi}^{\pi} 
\Omega(\hat{\mu}- \varphi) 
\cos   \left( \hat{\mu} - \varphi \right)  \, \rho(\varphi)  d\varphi.\\
$$
where $\rho(\varphi)$ represents the density of preferred phases of the tuning
curves. With $M$ equidistantly
spaced tuning curves along one dimension (with $x$ a scalar), we get
$\left\langle \hat{\kappa}\right\rangle = \left( 2 \pi  f_{max} \tau M \right) 
\kappa \exp(-\kappa) I_1(\kappa)$,  where $I_1$ is the modified Bessel function of
the first kind.
This  relates the expected concentration $\hat{\kappa}$ 
of the posterior probability to  the inverse tuning width $\kappa=\sigma^{-2}$ of the
tuning curves. Figure~\ref{fig:pmu_and_pk}  emphasizes the fact that  the variance 
of $\hat{\kappa}$ is large;  indeed, this is always the case, as the
$\text{variance}\!\left( \hat\kappa \right) \sim \kappa \langle \hat{\kappa}
\rangle$~\footnote{
For $u= \kappa  \sum_{j=1}^M n_j \cos( \,  \omega (x-\varphi_j) )$, one computes a
characteristic function
$\Phi_u (t) =\sum_{j=1}^M  \sum_{n_j=0}^\infty \exp( u(n_j)  t) p(n_j)= \exp\left\{
f_{max} M  \tau  e^{-\kappa}  {2 \pi} \cdot  (I_0(\kappa (t+1))  )- I_0(\kappa) )  \
\right\}
$. The characteristic function allows one to compute the moments of $P(u)$ (by taking
the $n$-th derivative of $\Phi_u(t)$ at $t=0$).
We then expand the  modified Bessel functions asymptotically to get the scaling
result in the text.
}.

\begin{figure}[!ht]
\begin{center}
\subfloat[][]{\label{subfig:posterior}
\includegraphics[angle=0,width=0.5\textwidth]
{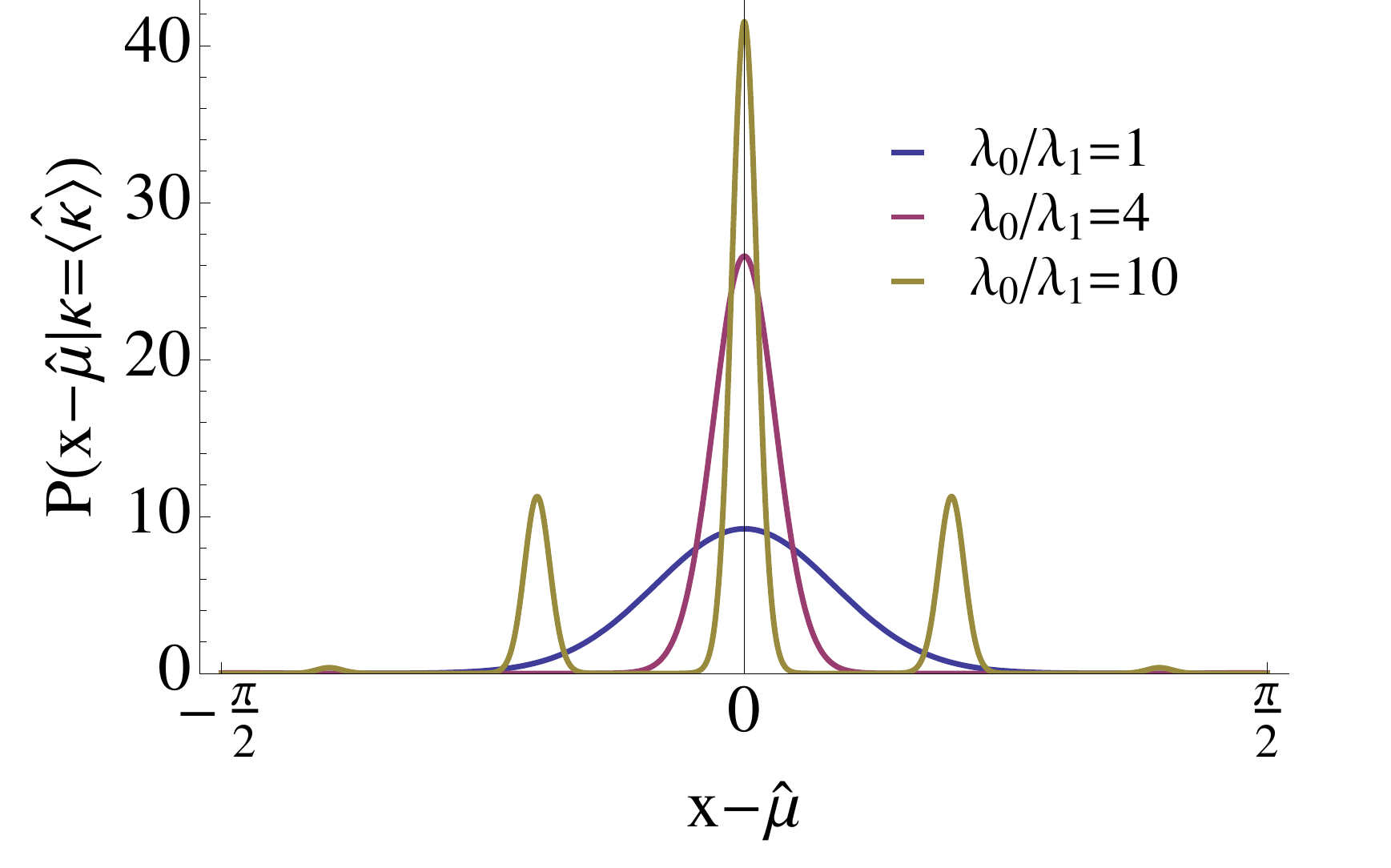} } 

\subfloat[][]{\label{subfig:two_mod}
\includegraphics[angle=0,width=0.5\textwidth]
{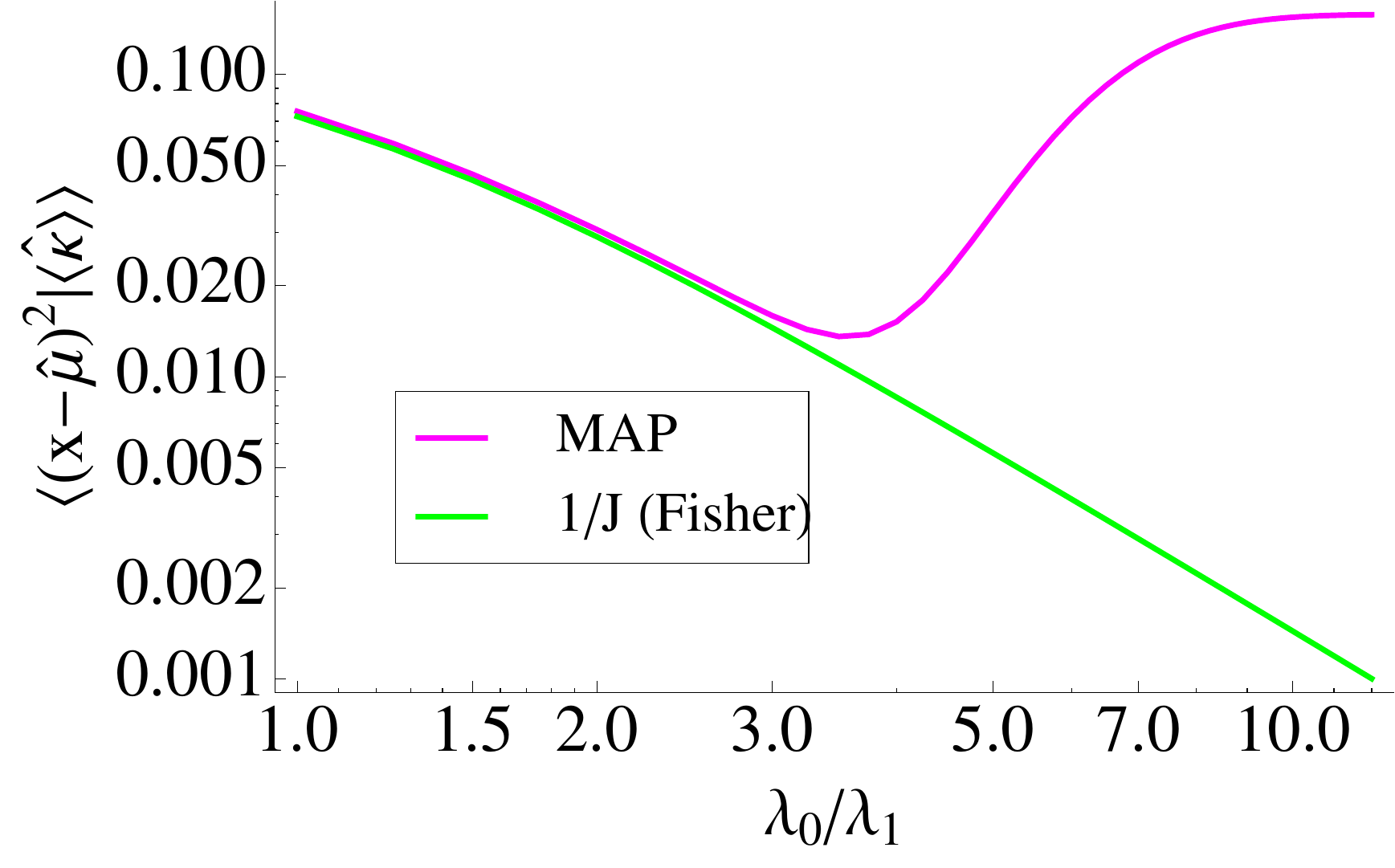}}
\end{center}\caption{The typical error in estimating $x$ as  the most likely stimulus
given the neuronal response in  a two-scale population code. To estimate this error, 
we replace  $\hat{\kappa}_i$ and $\hat{\mu}_i$ in
Eq.~\eqref{eq:multi_scale_posterior}
by their expected values. 
Without loss of generality, we consider $x=0$. The parameters
are $f_{\text{max}} \tau=1$, $M=16$, and $\sigma^2=1/2$.
In this simple approximation,  the  error in the maximum a posteriori (MAP) estimate
is
$ C^{-1} \int x^2 \exp\left\{\sum_{i=1}^L \left\langle \hat{\kappa}_i \right\rangle
\cos \left[ \frac{2 \pi x}{\lambda_i} \right] \right\}\, dx$,
with $C=  \int  \exp\left\{\sum_{i=1}^L \left\langle \hat{\kappa}_i \right\rangle
\cos \left[ \frac{2 \pi x}{\lambda_i} \right] \right\}\, dx$ normalizing the
posterior probability 
so that it integrates to unity.
When the scales $\lambda_0$ and $\lambda_1$ separate, the error initially
improves, since 
the posterior distribution narrows (Fig.~\ref{subfig:posterior}). At the same time, 
the secondary
peaks become increasingly more pronounced. 
 When the resolution limit of the module at the coarser scale $\lambda_0$ is reached,
no further refinement in the estimate of $x$ by the second module is possible.
Fig.~\ref{subfig:two_mod}
compares the error of the MAP estimate $\hat{x}$   to the asymptotic prediction
from the Fisher information. }
\label{fig:two_modules}
\end{figure}

Now consider a multi-scale population code consisting of $L$ modules, each with $M$
neurons, for a total  of  $N=L\cdot M$ neurons. Each module has a separate scale set
by the period 
$\lambda_i$, and within each module the angular preferences
are assumed to be equally spaced, i.e. \begin{equation} \varphi_i \in
\left\{0,\frac{2 \pi 
\lambda_i}{M
}, \ldots ,\frac{2 \pi (M-1) \lambda_i}{M}\right\}.\end{equation}

As described in ~\cite{Mathis2012b}, the spatial periods should obey:
\begin{equation}\lambda_{k+1} = \frac{S \cdot \lambda_k }{ \sqrt{J} },
\label{eq:nesting}
\end{equation}
with a  \textit{safety} factor $S\gg 1$ and $J$ being the Fisher information for the
module
at the coarsest scale. 
We can now use the posterior probability $P(x|\bm \Upsilon)$ to study how the typical
error in encoding $x$ depends on the ratio $\lambda_k/ \lambda_{k+1}$. For a
multi-scale
code, we have
\begin{equation}
P\left(x|\bm \Upsilon \right)= C'' \exp\left\{\sum_{i=1}^L \hat{\kappa}_i \cos \left[
\frac{2 \pi}{\lambda_i}  (x - \hat{\mu}_i)\right] \right\},
\label{eq:multi_scale_posterior}
\end{equation}
with $C''$ a new normalization constant.

 \begin{figure}[!ht]
\begin{center}
\includegraphics[angle=0,width=0.95\columnwidth]{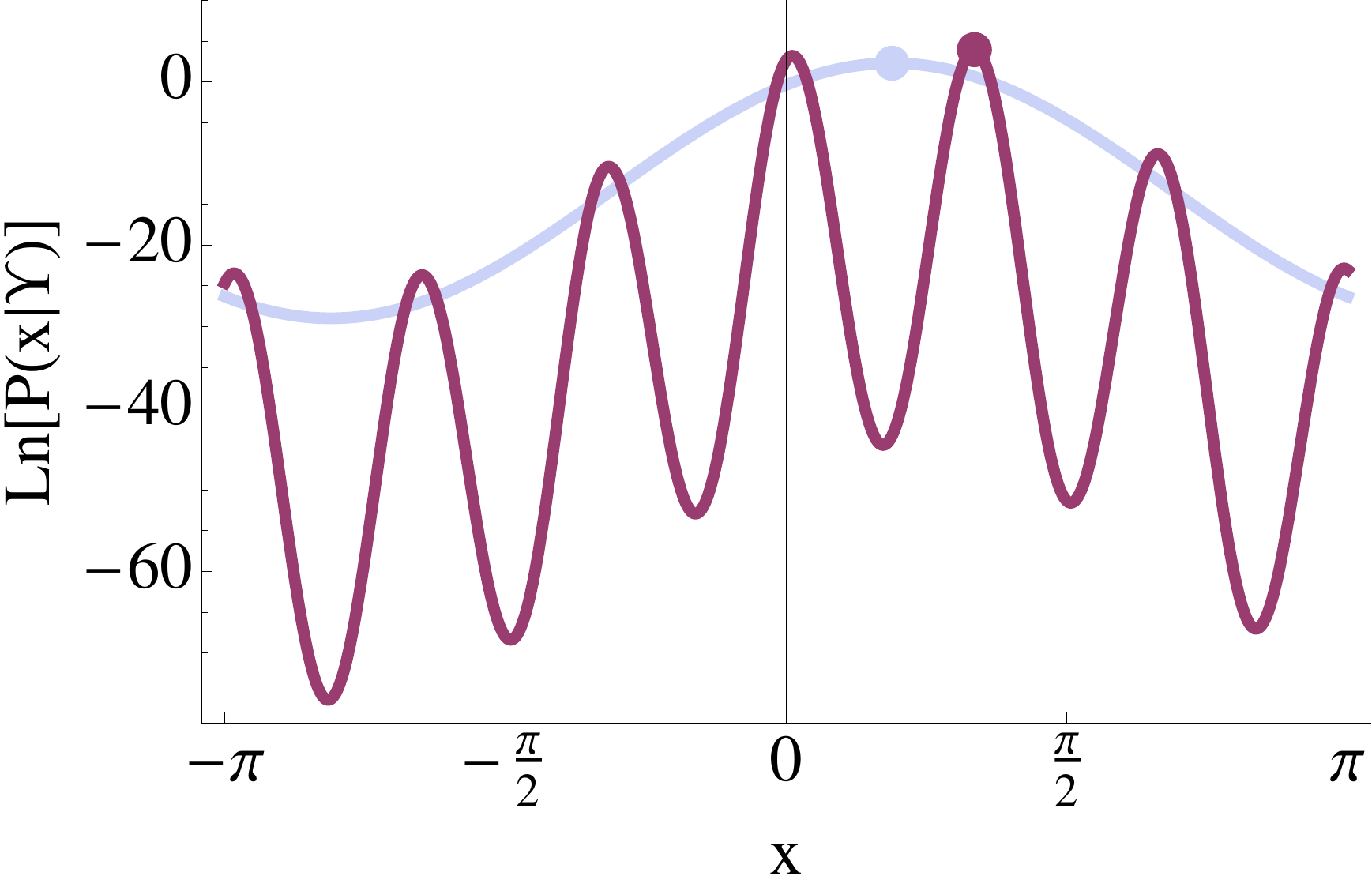}
\end{center}
\caption{For a given population response $\Upsilon$,  measured  when a stimulus $x=0$
is presented, the error of a multi-scale code can be worse than that of a
single-scale code. A particular population response gives rise to a posterior
probability $P(x|\Upsilon)$, an example of which is shown above for a population code
with  $M=8$, $f_{max} \tau=1$ and $\sigma^2=0.86$. The maximum likelihood estimate 
$\hat{x}$ in this example is 0.59 for a single module (light blue), and 1.05 for two
modules (pink)--these are indicated by dots marking the highest peaks, respectively.
Compared to a single module, the expected error $\left\langle
(x-\hat{x})^2\right\rangle$  is five times larger when using two modules. }
\label{fig:catastrophic_error}
\end{figure}

Consider a two-scale population code, as in Fig.~\ref{fig:two_modules}.  As a first
approximation to the typical (or median) error, set
$\hat{\mu}_i$ and $\hat{\kappa}_i$ to their expected values in
Eq.~\ref{eq:multi_scale_posterior}; later, we will refine the numerical computation
to reflect the true average error.
As $\lambda_1$ is made smaller relative to $\lambda_0$, the typical  error
improves initially, but then worsens as  $\lambda_1$ falls below the resolution of
the module with length scale  $\lambda_0$. For $\lambda_1 \ll \lambda_0$,
$P\left(x-\hat{\mu}\right)$ develops side peaks at integer multiples of $2
\pi/\lambda_1$ (Fig.~\ref{fig:two_modules}a); the expected typical error worsens.
In the example shown, $1/\sqrt{J} \approx 16.6$, so we deduce that the Cram\'er-Rao
bound of Eq.~\eqref{eq:cramer-rao} can be attained for 
safety factors $S > 5$ (Fig.~\ref{fig:two_modules}b). For $\lambda_0/\lambda_1 \to
\infty$, the resolution of the population code with two modules
 reverts to that of the single (coarse-scale) module.

\begin{figure}
\begin{center}
\subfloat[][]{\label{subfig:3D_S6}
\includegraphics[angle=0,width=0.24\textwidth]
{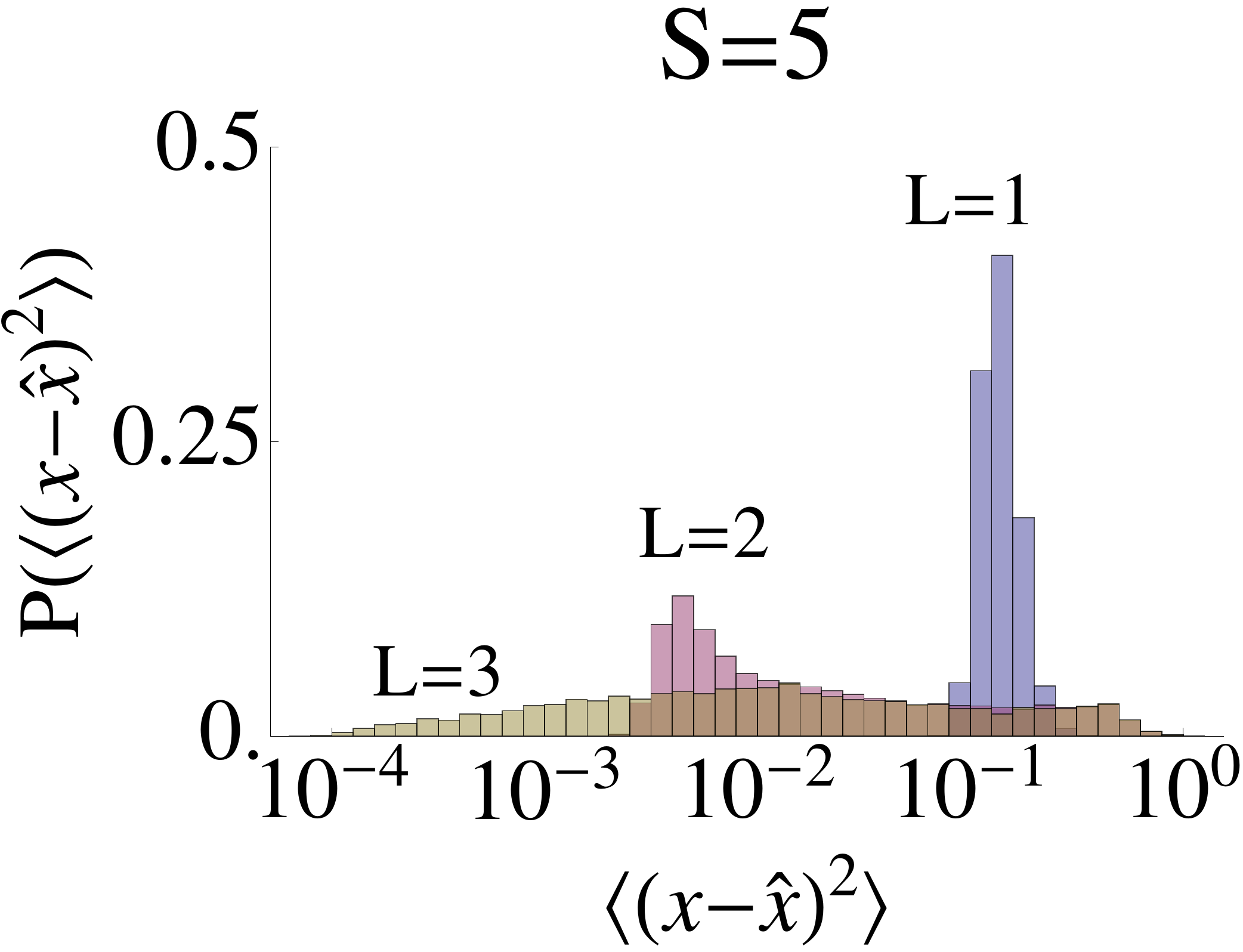} } 
\subfloat[][]{\label{subfig:3D_S5}
\includegraphics[angle=0,width=0.24\textwidth]
{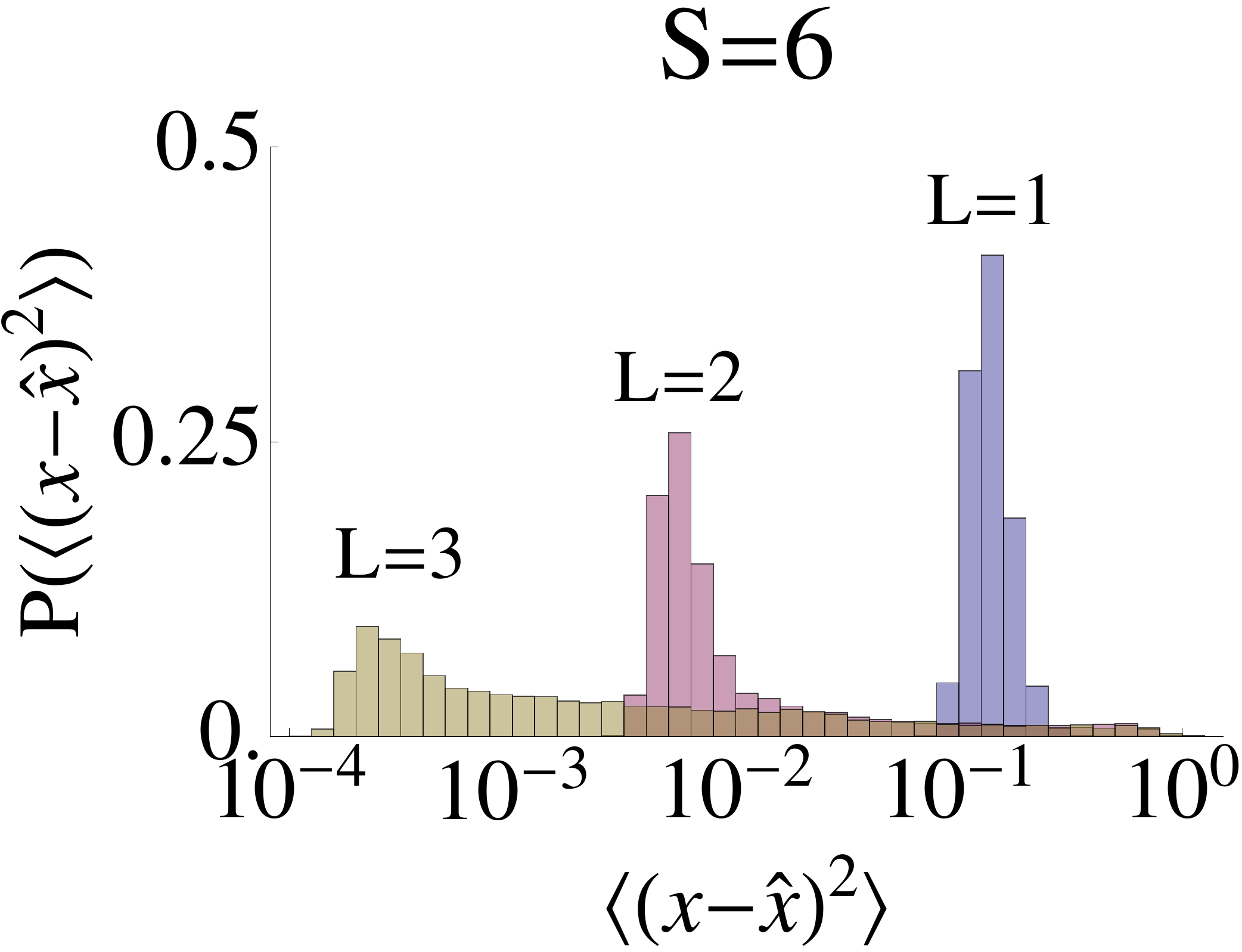}} \\
\subfloat[][]{\label{subfig:kappa_correlation}
\includegraphics[angle=0,width=0.23\textwidth]
{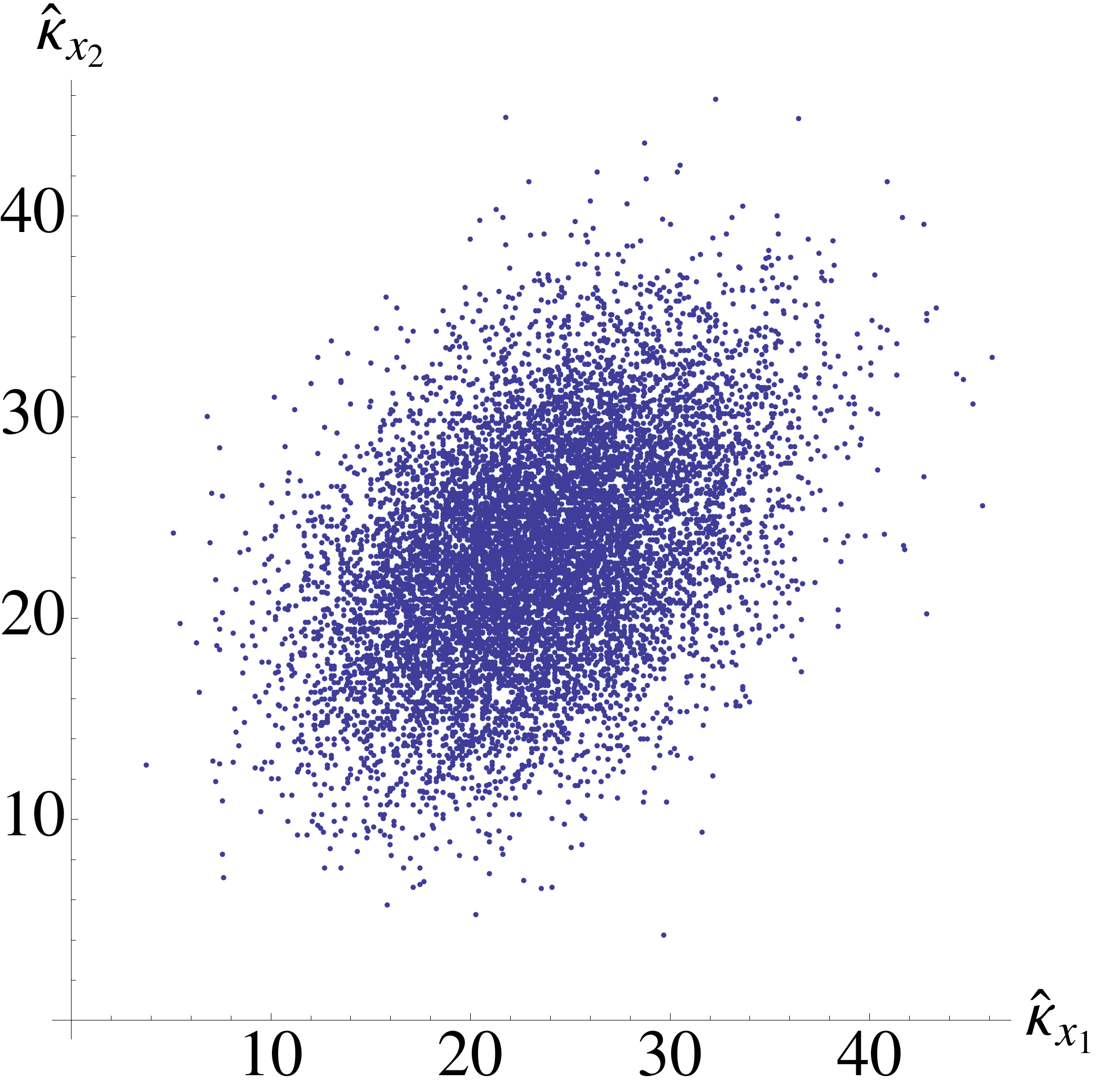}} 
\subfloat[][]{\label{subfig:mu_correlation}
\includegraphics[angle=0,width=0.23\textwidth]
{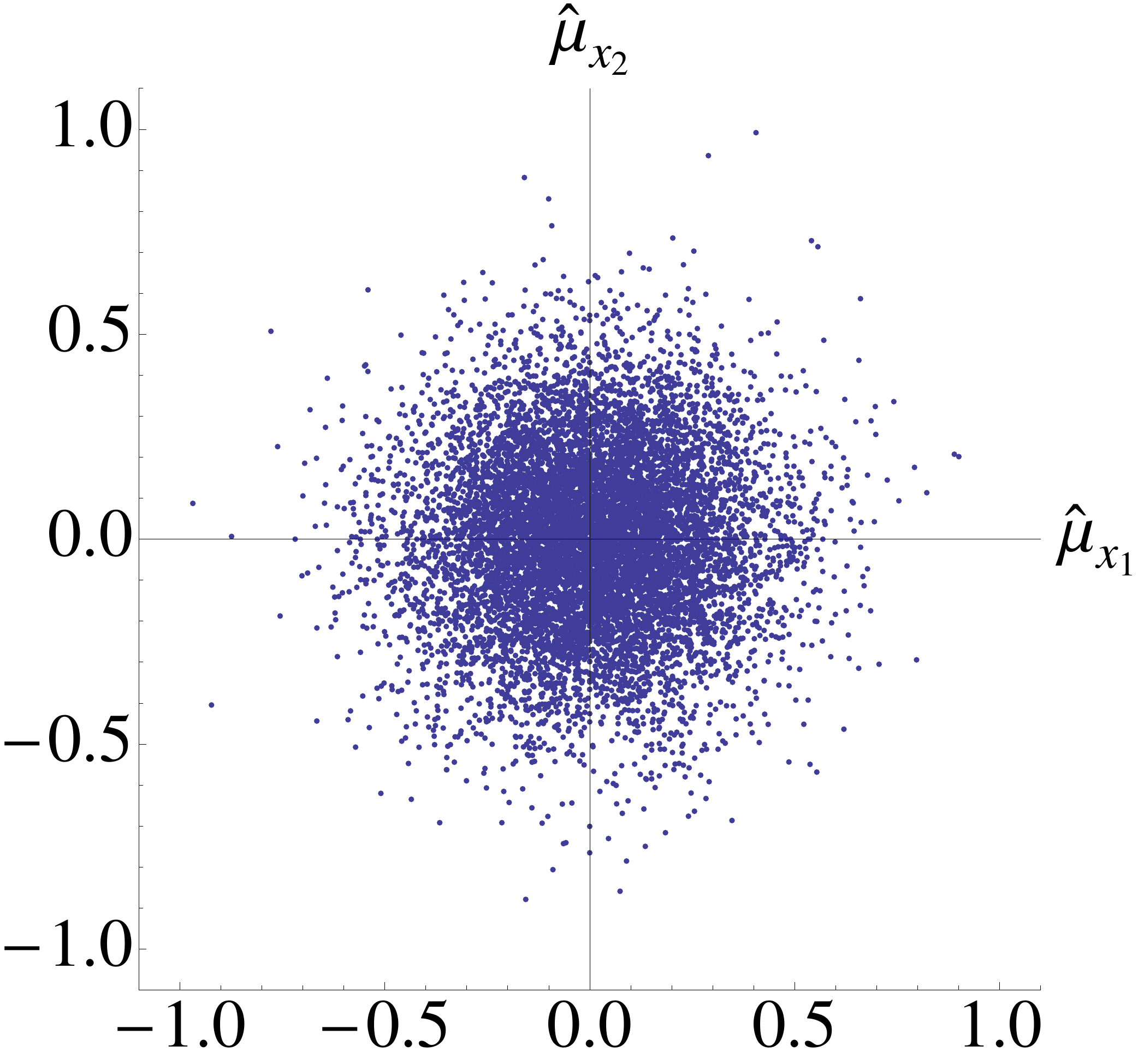}}
\end{center}\caption{Multiple modules can refine the estimate of $x$, decreasing the
error by more than an order of magnitude
for each new module added (Fig.~\ref{subfig:3D_S6}). The maximum likelihood
estimate of $\hat{\bm x}  \in [-\pi,\pi)^3$ and the posterior distribution 
$P((x-\bm x)|\bm  \Upsilon)$ are evaluated numerically for each sampled population
response $ \bm \Upsilon$. The top panels show the
distribution of expected errors for population codes with $L=1,2,$ and 3 modules
encoding $\bm x$ at different spatial scales. Every module contains  $M=8^3=512$
neurons with tuning curves that are equidistantly spaced.  The relative tuning width
is $\sigma^2=0.86$ for each neuron, and Eq.~\eqref{eq:nesting} dictates the geometric
progression
of spatial scales. When $S=5$,  the refinement scheme breaks down by the third module
($L=3$, Fig.~\ref{subfig:3D_S5}). The errors have been sampled 10,000 times. For a
multi-dimensional stimulus $\bm x$, the posterior
probability $P(\bm x |
\Upsilon)$  depends on the random variables $\hat{\kappa}_{x_\alpha}$ and 
$\hat{\mu}_{x_\alpha}$along the
different dimensions.
The random variables $\hat{\kappa}_{x_\alpha}$  are correlated
(Fig.~\ref{subfig:kappa_correlation}), with a pairwise correlation of
0.44. This correlation depends on $\sigma^2$ and $f_{\max} \tau$, and can be stronger
or weaker. In contrast, the random variables  $\hat{\mu}_{x_\alpha}$ are uncorrelated
(Fig.~\ref{subfig:mu_correlation}).}
\label{fig:high_dimensional}
\end{figure}

Going beyond the first-order approximation, we can sample  the network's response
$\bm \Upsilon$ repeatedly,
and thereby sample $\hat{\kappa}_i$ and $\hat{\mu}_i$. Numerically, we can thus
estimate the average error
 of the population code from Eq.~\eqref{eq:multi_scale_posterior} without resorting
to averaging $\hat{\kappa}_i$,
 even for multi-dimensional stimuli with $D > 1$. 
 We proceed in three steps: first, we numerically determine the maximum a posteriori
estimate $\hat{\bm x}$ of ${\bm x}$ by maximizing the argument in the exponential
of
Eq.~\eqref{eq:multi_scale_posterior}; secondly, we integrate over the posterior
distribution to obtain  the expected error 
  ${\left\langle \left( \bm x - \bm \hat{\bm x} \right)^2 \right\rangle}_{P(\bm x|
\Upsilon)}$ for each response $\bm \Upsilon$; lastly,
 we build a histogram of the expected errors.
 
 The posterior probability in Eq.~\eqref{eq:multi_scale_posterior}  depends on the
superposition of oscillatory functions in the argument to the exponential function.
Decoding the population response $\Upsilon$ 
 can lead to combinations of $\hat{\mu}_i$'s and $\hat{\kappa}_i$'s that cause the
oscillatory functions to interfere constructively, but at the wrong location; witness
 Fig.~\ref{fig:catastrophic_error}, in which the MAP estimate of $x$ actually becomes
{\em worse} when using a second module to refine the estimate
 from the first one. 

For multidimensional stimuli $\bm x$ with $D > 1$, the risk of catastrophic error is
cumulative, as each new dimension
adds a new possibility to make a decoding mistake.  As the dimension $D$  or the
number of modules $L$ increases,
the requirement that $S \gg 1$ must be made more stringent.
 
We numerically analyze a multi-scale population code  for a stimulus $\bm x$  of
dimension $D=3$, so that  $\bm x$ is contained in the normalized
interval $[-\pi,\pi)^D$. The network has $M=8^D$ neurons with tuning curves that are
the product of  one-dimensional {\it von Mises} functions
$\Omega_i(\bm x) = \prod_{\alpha=1}^D \Omega_i(x_\alpha)$. For $D \geq 3$, there is
an optimal $\sigma^2$ for the tuning curve
that maximizes the Fisher information~\cite{Mathis2012a}; we use the value of
$\sigma^2 \approx 0.86$ appropriate for $D=3$. 
If one takes a marginal safety factor $S=6$ in Eq.~\eqref{eq:nesting}, the errors
decrease by more than an order of magnitude for each new module added,
as predicted by the Fisher information's scaling  as $J^L$
(Fig.~\ref{fig:high_dimensional}a); adding neurons to a single module (at a single
spatial scale), in contrast,  leads only
to an improvement that is linear in $N$. As one decreases the safety factor to $S=5$,
catastrophic errors accumulate; Fig.~\ref{fig:high_dimensional}b shows that the third
module does not
improve the average error. 
Although the posterior probability factorizes in the $D$ dimensions, the components
of $\hat{\kappa}$ along these dimensions are correlated
(Fig.~\ref{subfig:kappa_correlation}).
Hence, decoding a population response that is uncertain in $x_1$, say, will also
likely be uncertain in $x_2$ and $x_3$. As for the single-scale population code (see
Fig.~\ref{fig:pmu_and_pk}), the expected error as a function of the response
$\Upsilon$ can be greater or less than the inverse Fisher information $1/J$. On
average, though, the 
expected error of the MAP estimate is bounded from below by $1/J$, and for $S >5$ the
bound is close to the average error.

\section{Population coding model with noise correlations}

Until now, we have considered a network of $N$ neurons  in which the fluctuations in
neuronal activity depend
only on the stimulus $x$, not on the activity of other neurons.
We now treat the more general case of neuronal activity that has  non-trivial
\textit{correlation structure}.
Correlations might adversely affect the Fisher information and potentially even make
a multi-scale code for cortex infeasible.

Before estimating the noise correlations for real spike trains from grid cells
recorded in entorhinal cortex,
let us extend the standard model for ensembles of correlated cells with unimodal
tuning
curves~\cite{Shamir2006,Ecker2011} to grid codes. To keep the analysis simple, 
we consider only a one-dimensional stimulus $x \in [-\pi,\pi)$, and compare the
Fisher information to 
the mean squared error.

\subsection{Model of noise correlations}

In the equation for  the response of neuron $i$, $\Upsilon_i(x) = \Omega_i(x) +
\eta_i(x)$, let $\eta_i(x)$ now 
follow a multivariate normal distribution with zero mean and a non-diagonal
covariance matrix
$Q(x)$, which implies that the neurons are no longer statistically independent. 
More specifically, let us posit the model in Ref.~\cite{Ecker2011}, for which the
covariance matrix is given by the product
\begin{equation}
 Q_{ij}(x) = \sqrt{\Omega_i(x)} \cdot r_{ij} \cdot \sqrt{\Omega_j(x)}.
 \label{covariance}
\end{equation}
This model assumes that the correlation factor between two neurons $r_{ij}$ is
independent
of the stimulus $x$, hence $Q_{ij}(x)$ quantifies the "noise correlations". In the
limit in which neurons
become statistically independent, $r_{ij} = \delta_{ij}$, $Q_{ij}(x) = \Omega_i(x)
\delta_{ij}$; 
in other words, the variance scales with the mean response $\Omega_i(x)$, just as in
the discrete Poisson model.

To complete the model, we need to determine the  correlation coefficients $r_{ij}$.
Within a  module, each cell's tuning curve has a spatial phase $\varphi_i$. The
functional
organization of 
cortex~\cite{Ecker2011} suggests that cells with similar coding properties will have
larger correlation coefficients;
this is, indeed, the case for grid cells in cortex, as we will show in detail later. 
Therefore, 
we let the correlation coefficient between two cells depend on the difference $d$ in
spatial phases $\varphi_i$ and $\varphi_j$:

\begin{eqnarray}
r_{ij} & = &c\left( -\pi + (\varphi_i-\varphi_j+\pi) \!\!\! \mod 2\pi)\right)
\nonumber \\ & & + 
\delta_{ij} (1-c(0)).\label{corrcoeffdef}
\end{eqnarray}

Here  $c$ is a monotonically decreasing
function. We will use $c(d) = c_0  \displaystyle \cdot \exp
\left(-\frac{d}{\nu} \right),$ with $\nu=1$.  Across modules the correlations are
assumed to vanish.

\subsection{Fisher information for correlated populations}

For the  model of Eq.~\eqref{covariance}-\eqref{corrcoeffdef}, the Fisher information
can be written as a
sum~\cite{Kay1993,Shamir2006,Ecker2011}: \begin{equation}
J(x) = J_{mean}(x) + J_{cov}(x)
\end{equation}
with the following individual parts:
\begin{eqnarray}
& J_{mean}(x) = (\Omega'(x))^t Q(x)^{-1} \Omega'(x)
\\
& J_{cov}(x) = \frac{1}{2} \mathrm{Tr}\left( \left(Q'(x)
Q(x)^{-1} \right)^2\right).
\end{eqnarray}

In these equations, $\Omega'$ and $Q'$ are the derivatives with respect to the
stimulus variable
$x$.  $J_{mean}$ depends on
the changes of the mean firing rate $\Omega'$ and $J_{cov}$ depends on
changes in the covariance structure $Q'$.

We will compare the Cram\'er-Rao estimate of the error based on the Fisher
information to the  least mean square estimator (MSE) for the neural population code.
The latter
is given by \begin{equation} \widehat{x_{MSE}}(K) = \frac{\int_{-\pi}^{\pi} x
P(K|x) dx}{\int_{-\pi}^{\pi} P(K|x) dx}. \end{equation} Numerically, we divided the
stimulus space $[-\pi, \pi)$ into $m=10^5$ equidistant points
$\{x_0,x_1, \ldots, x_m \}$ and computed the MSE
for  $n=15 ,000$ uniformly distributed positions.  After averaging over the
squared residues, one obtains 
the mean square (estimate) error:\begin{equation} \epsilon_{MSE}^2 = 1/n \sum_k
(\widehat{x_{MSE}}(K(x_k))) - x_k) ^2.      \end{equation}

\ifdraft
\subsection{Correlations in multi- and single-scale population codes}
\else
\subsection{Correlations in multi-\\ and single-scale population codes}
\fi

We studied how the Fisher information $J$ depended on the population size $N$ and the
correlation peak $c_0$, either
by increasing the number of modules $L$ or by adding multiples of $M$ neurons to a
single module.
For the simulations, we set the peak rate  to $f_{max}=20$ Hz and the tuning
width to $\sigma^2=1/2$. Qualitatively these choices are not crucial, as long
as there are enough neurons to cover the space given $\sigma$ and the peak spike
count is larger than one (cf. Ref~\cite{Mathis2012a}).

\begin{figure}[!ht]
\centering
\includegraphics[angle=-90,width=0.5\textwidth]{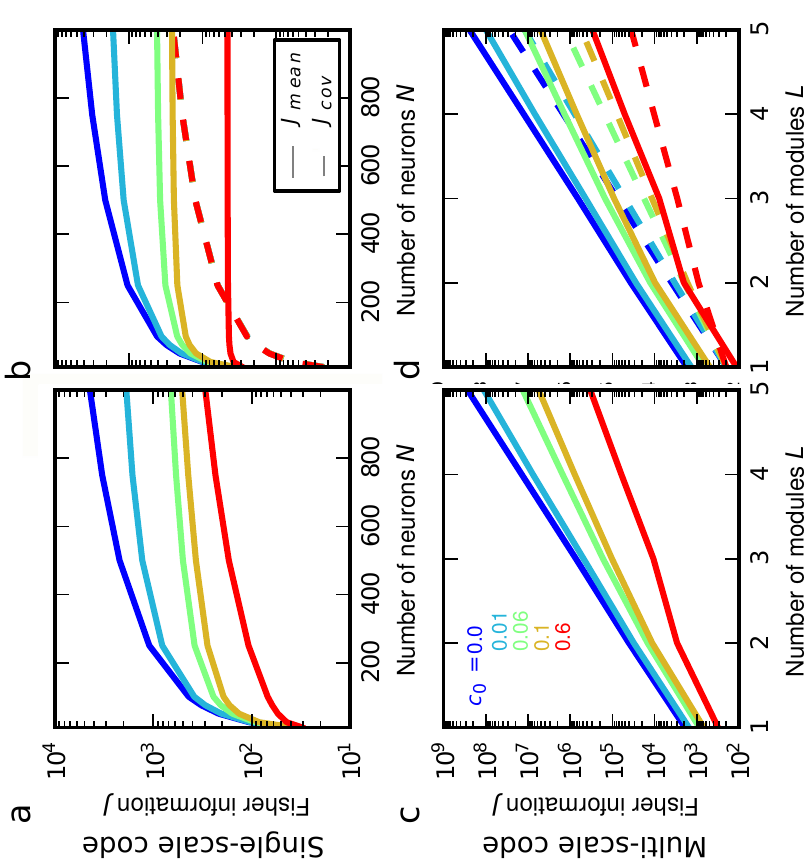} 
\caption{Fisher information for population codes with correlations. We evaluated the
Fisher information at position $x=0$.
\textbf{a}: The
total Fisher information $J$ for a population of $N$ place cells with
correlation peak $c_0$. For zero-correlation the Fisher information grows
linearly in $N$. For larger correlation coefficients the
Fisher information falls, but eventually grows linearly in $N$, as indicated by
considering the two components of $J$ individually, see subfigure b. 
\textbf{b}: The same simulation, but the two parts of the Fisher
information $J_{mean}$ and $J_{cov}$ are shown separately in solid and dashed
lines, respectively. The mean term saturates for increasing correlation peak $c_0$,
but the covariance term grows linearly and is in fact independent of the correlation
peak $c_0$. \textbf{c}: Fisher information for grid code without
inter-module correlation. The total Fisher information $J$ for a
population of $L$ modules and correlation peak $c_0$. Each module contains $M=200$
neurons. Even for increasing correlation, the population Fisher information still
grows stronger than linearly. The stronger the correlation coefficient becomes
the smaller the contraction factor $C/\sqrt{J}$ becomes, and therefore
the smaller the growth. \textbf{d}:  The same simulation, but the two parts of the
Fisher information $J_{mean}$ and $J_{cov}$, are shown separately in solid and dashed
lines, respectively. }\label{fig:Correlations}
\end{figure}

In the absence of noise correlations, the Fisher information of
a single module grows linearly in $N$~\cite{Zhang1999,Mathis2012b}. For
rising correlation amplitude $c_0$, the Fisher information decreases, yet still grows
linearly with $N$ (Figure~\ref{fig:Correlations}a). This effect can be
explained by considering the two components $J_{mean}$ and $J_{cov}$ individually, as
shown in Fig.~\ref{fig:Correlations}b. While the former saturates, the latter grows
linearly in $N$, independently of the degree of  correlation. This result is well
known, e.g., see Shamir and Sompolinsky~\cite{Shamir2006,Ecker2011}.

For a nested, multi-scale population code,                                          
the Fisher information in each module decreases as the  peak intra-module correlation
 $c_0$ 
increases. For $c_0=0$,  the spatial periods should obey the previously derived
relationship
for the population code to attain the Cram\'er-Rao bound:
\begin{equation}\lambda_{k+1} = \frac{S \cdot \lambda_k }{ \sqrt{J} },
\tag{\ref{eq:nesting}}
\end{equation}
with \textit{safety} factor $S \gg 1$ and $J$ the Fisher information of the first
module (at the coarsest scale, see~\cite{Mathis2012b}, Eq. 8 and discussion
thereof). 
The same still holds true for $c_0 > 0$, only the factor $\sqrt{J}$ is less.
Thus, the population Fisher information of a grid code, despite still growing
exponentially, grows more slowly in $N$ for rising $c_0$.
Figures~\ref{fig:Correlations}c~and~d depict the Fisher information of a grid code
with up to $5$ modules and
$M=200$ neurons per module.

\begin{figure}[ht]
\centering
\includegraphics[angle=-90,width=0.45\textwidth]{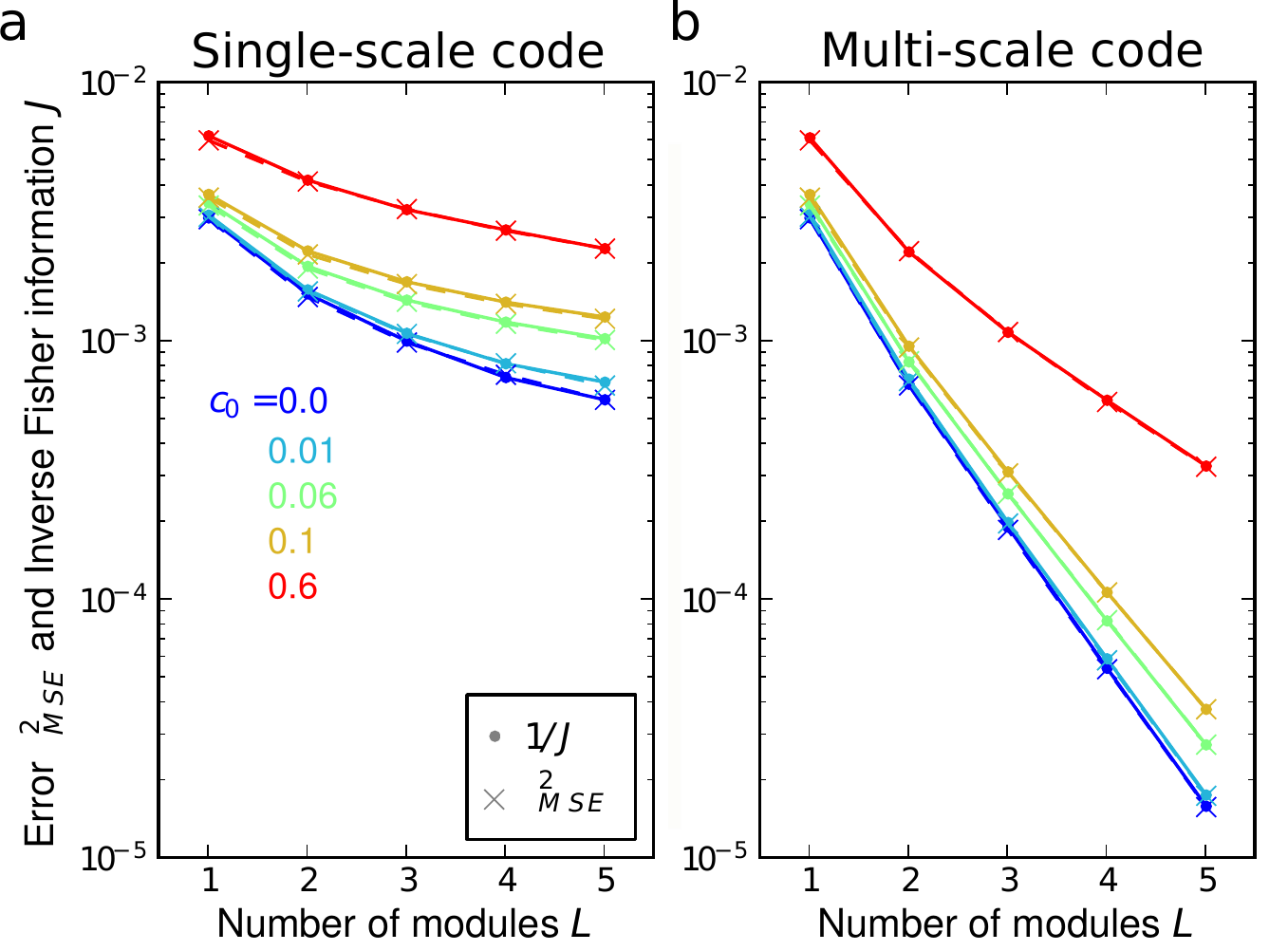}
\caption{Inverse Fisher information $1/J$ and $\epsilon_{MSE}^2$ for an
ensemble of cells with tuning curves on a single scale (left) and on multiple scales
(right), for 
varying degrees of noise correlation between neurons in a module. 
Each ensemble contains $L\cdot M$ equi-distantly arranged
neurons, with $M=64$, $f_{max}=10\, \text{Hz}$, and $\sigma^2=1/2$. For a pair of
neurons, the noise correlation
depends on the difference of their preferred phases,  according
to Eq.~\eqref{corrcoeffdef},  with $\nu = 0.19$ and varying $c_0$. There are no noise
correlations 
between modules.
a: each module contains tuning curves of period $\lambda_0$.
b: the periods are staggered according to Eq.~\eqref{eq:nesting} with
$S=10$.
 For the parameters  considered here, the MSE is close to the Cram\'er-Rao bound.}
\label{fig:PcvsGC_intra}
\end{figure}

Under certain conditions the Cram\'er-Rao bound is not
tight~\cite{Bethge2002,Berens2011,Mathis2012a}, so we corroborated our
results by computing the mean square error (MSE) for these population codes.
Figure~\ref{fig:PcvsGC_intra} shows that the MSE is close to the inverse Fisher
information for this set of parameters.

\section{Noise correlations of biological grid cells}

\subsection{Estimation of spatial period, phase and noise correlation in pairs of
grid cells}
\begin{figure}[!ht]
\begin{center}
\includegraphics[angle=0,width=0.9\columnwidth]
{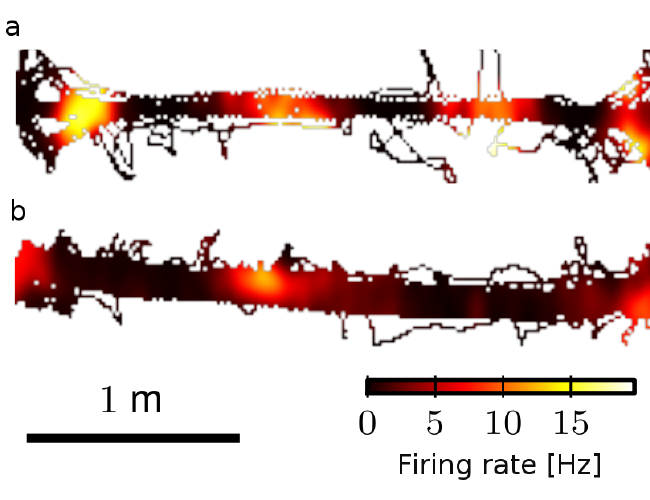}
\end{center}\caption{Firing rate maps $\Omega_i(x,y)$ for two neurons in the data set
from Hafting {\it et al.}~\cite{Hafting2008}, 
in which rats ran back and forth on a linear track. We filtered the spikes as
described in Eq.~\eqref{Kernelfiltering}.  Cells in \textit{a} and \textit{b}  differ
in their
spatial periods, which we estimated to be $88$ and $177$ cm, respectively, from the 
first peak in the autocorrelation.  Prior to the runs on the track, these cells were
recorded while the rat foraged in a two-dimensional enclosure, which revealed that
the cells had a hexagonal map
of spatial firing.  It is not guaranteed, however, that the track is aligned to one
of the principal axes of the hexagonal map's lattice, so not all  cells have a linear
track firing rate 
$\Omega_i(x,y)$ that is perfectly periodic (see also
Ref.~\cite{Domnisoru2013,Kempf2012}.}
\label{fig:GridCell}
\end{figure}

To estimate the correlations between in a real neural population with multi-scale
coding properties, 
we analyzed grid cell data recorded by Hafting et al.
(\cite{Hafting2008}, available at
\url{http://www.ntnu.no/cbm/moser/gridcell}). As the experimental methods are
extensively covered in the original publication~\cite{Hafting2008},  we briefly
summarize the  details relevant to our study here. In the experiments, rats ran back
and forth 
 for ten minutes on a  $320$ cm long and $10$ cm wide linear track\footnote{For a few
cells, multiple $10 \, \text{min}$ long sessions were recorded, but we only
used  the first $10 \, \text{min}$ for each cell for this analysis};  during each
run, the
trajectory was  tracked using a head-fixed light emitting diode, and the neuronal
activity in the
medial entorhinal cortex (mEC) was recorded using extracellular tetrodes. From these
signals, spike times of several single cells in mEC were isolated. A subset of these
cells fired
at multiple, periodically spaced locations on the linear track, separated by
stretches of the track on which the cells did not fire (Fig.~\ref{fig:GridCell}).
These cells are called 'grid cells' and exhibit different spatial periods in their
firing rate maps. 
The spatial firing of a grid cell  is  characterized by its spatial period (i.e.
average peak-to-peak distance of firing
fields) and phase (position of the first 
peak, for instance, relative to a reference point.)~\cite{Hafting2005}. 
Neighboring cells in mEC  tend to have a  similar spatial
period in their firing pattern, but differ in their spatial
phases~\cite{Hafting2005}. 


 Overall, the data set contained $97$ cells.
Left- and rightward runs were treated separately, as the cell's firing pattern for
the two directions was often
different (see, for instance, Fig.~\ref{fig:GridCell}a). As is common
practice~\cite{Hafting2008,Brun2008}, we excluded
the first $30 \, \text{cm}$ on both sides of the linear track from consideration;
here the rats slow down and turn around.
 For pairs of grid cells  that were recorded at the same time and in the same
animal, we estimated the noise correlations and the phase difference
 between the firing patterns.  $302$ such pairs were analyzed. The phase difference
of two periodic signals only exists if their frequencies are similar. The spatial
period of each grid cell must  be estimated
from  spike trains that are variable from run to run, so we proceed as follows: 
\begin{itemize}
\item[(i)] we determine the firing
rate for each cell by Gaussian kernel filtering in the spatial domain:
\begin{equation}
\Omega(x,y) = \frac{\sum_{s}
\exp{\left(-\frac{(s_x-x)^{2}}{2\sigma_x^{2}}-\frac{(s_y-y)^{2}}{2\sigma_y^{2}}
\right)}}{dT \sum_{t} 
\exp{\left(-\frac{(\gamma_t(x)-x)^{2}}{2\sigma_x^{2}}-\frac{(\gamma_t(y)-y)^{2}}{
2\sigma_y^ { 2 } }\right)}} \label{Kernelfiltering}
\end{equation}
where $s=(s_x,s_y) \in S$ are the spike positions.
$\gamma_t$ is the discretized trajectory, sampled in  $dT=0.02 \, \text{s}$
steps.
 We used $\sigma_x=3\,\text{cm}$ and
$\sigma_y=3 \, \text{cm}$. The map of Eq.~\eqref{Kernelfiltering}  was computed on a
discretized grid with $N_x \times N_y = 160
\times 10$ bins denoted by $(x_k,y_l)_{\{1 \leq k \leq N_x,1 \leq l \leq N_y\}}$
(see Fig.~\ref{fig:GridCell}b). Then we averaged along the $y$ axis and obtained the
firing rate profile $\Omega_i(x_k)$ for $1\leq k \leq N_x$ and each cell
$i$. 
\item[(ii)] The spatial period $\lambda_i$ is defined as the  first peak in the
autocorrelogram of the firing map $\Omega_i(x_k)$.
\item[ (iii)] For each cell pair $(i,j)$
we compute the cross-correlogram of $\Omega_i(x_k)$ and $\Omega_j(x_k)$, and the
spatial period $\lambda_{ij}$ as the first peak in the cross-correlogram.
\item[(iv)] If
$\lambda_{ij}$ differs by maximally $20\%$ from both
$\lambda_i$ and $\lambda_j$, we assume that the cells are from the same module
(i.e. share the spatial period) and define their phase difference
$\widehat{\varphi_{ij}}$ as the position of the peak in the cross-correlogram modulo
$\lambda_{ij}$. 
\item[(v)] Then we define the relative phase difference $\varphi_{ij} = 2\pi
\cdot \widehat{\varphi_{ij}} / \lambda_{ij}-\pi \in [-\pi,\pi]$.
\end{itemize}

For each pair $i,j$, we compute the noise correlations as follows:
\begin{itemize}
\item[(i)] From the spike
times$ t_k^{(j)}$ of each neuron $j$ we compute the temporal firing rate by Gaussian
kernel
filtering: 
\begin{equation}
f_j(t) = \frac{1}{\sqrt{2\pi} \sigma} \sum_{k}
\exp{\left(-\frac{(t_k^{(j)}-t)^{2}}{2\sigma^{2}} \right)}
\end{equation}
We used $\sigma=20\,\text{ms}$ and evaluated the firing rate on a $1\,\text{ms}$
fine temporal grid. 
\item[(ii)] We discretize the environment in $N_x \times N_y = 160
\times 10$ bins denoted by $(x_k,y_l)_{\{1 \leq k \leq N_x,1 \leq l \leq N_y\}}$.
For each session we compute the entry and exit times into these bins. Thus for
each bin $(k,l)$ we
get a list of $S_{k,l}$ trajectory segments $\gamma_{k,l}^s$ denoting the s-th path
trough the bin with entry time $\alpha_{k,l}^s$ and exit time $\omega_{k,l}^s$. 
\item[(iii)]
We compute the average firing rate for each cell $j$ and path
segment $s$: \begin{equation}
\overline{F_{k,l}^{j,s}}=\frac{\int_{\alpha_{k,l}^s}^{\omega_{k,l}^s} f_j(t)
dt}{\omega_{k,l}^s-\alpha_{k,l}^s}. \end{equation}
\item[ (iv)] These
rates allow us to compute the noise correlation of cell $i$ and $j$ in each bin
$(k,l)$ by computing the correlated response fluctuations around the means:
\begin{equation}
 c^{i,j}_{k,l} =\frac{\mathbb{C}ov_s \left(
\overline{F_{k,l}^{j,s}},\overline{F_{k,l}^{i,s}} \right)}{\sqrt{
\mathbb{V}_s \left(\overline{F_{k,l}^{i,s}}\right)
\cdot \mathbb{V}_s \left(\overline{F_{k,l}^{j,s}}\right)}}.
 \end{equation} 
 \item[(v)] These values per bin are then averaged over all bins to get the
noise correlations of cell $i$ and $j$:
\begin{equation}
 c_{i,j}= \frac{1}{N_x N_y} \sum_{k,l} c^{i,j}_{k,l}.
\end{equation}
\end{itemize}

\subsection{Estimated noise correlations of grid cells}

\begin{figure}[!ht]
\begin{center}
\ifdraft
\includegraphics[angle=0,width=0.6\columnwidth]{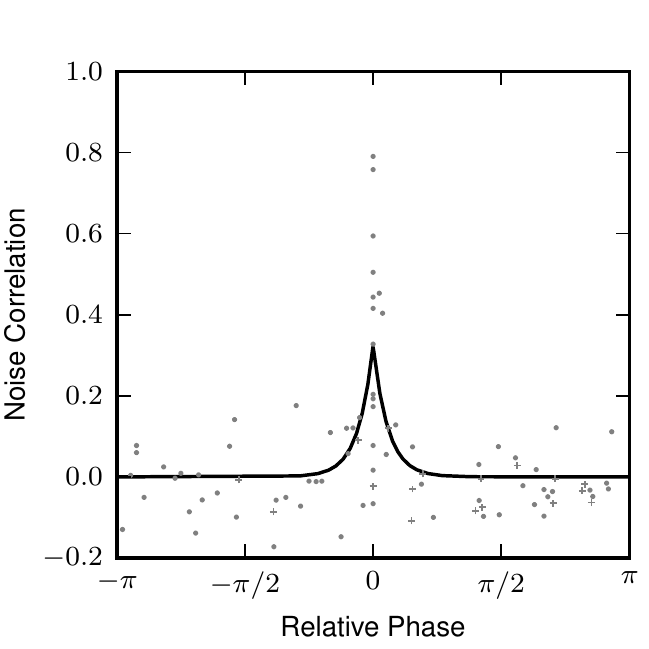}
\else
\includegraphics[angle=0,width=\columnwidth]{Fig9.pdf}
\fi
\end{center}\caption{Noise correlation $c_{i,j}$ vs. relative phases $\varphi_{i,j}$
of
$87$ pairs of grid cells from the same experimental session as described in the
main text. Gray dots indicate pairs from different tetrodes and gray crosses from the
same tetrode, respectively. Black continuous line
is a least-mean-square fit of the function $c_0 \exp(-|\varphi|/\nu)$ with values
$c_0=0.32$ and $\nu = 0.18$.}
\label{fig:NCvsrelsigma}
\end{figure}

In $87$ of $302$ grid cell pairs, both neurons belonged to the same module; i.e.,
they shared a common spatial period. We computed the noise correlations and
relative phases for these $87$ pairs (in $15$ pairs, both neurons were recorded on
the same tetrode). The relationship of noise correlation to phase
difference is shown in Fig.~ ~\ref{fig:NCvsrelsigma}. A least mean squares fit  of 
$c_0 \exp(-|\varphi|/\nu)$
to the data  yielded $c_0=0.32$ and $\nu = 0.18$. 
There are $62$ pairs with dissimilar phases (defined as $|\varphi|>0.36=2 \nu)$).
This group
has a mean noise correlation value of  $-0.021 \pm 0.0006$ (mean $\pm$ standard error
of mean). These neurons are, hence, uncorrelated. For similar phases, the maximal
noise correlation reaches $0.8$; on average, it  is $0.24 \pm 0.01$ for the other
$25$ pairs. These data indicate that  the noise correlations fall the farther apart
the spatial phases are.

\section{Conclusion}

A common objection raised to using the Fisher information is that  it only provides a
bound
to the resolution of the population code.
When the response is discrete or sampled for short durations, this bound is not
attainable by maximum
likelihood decoding~\cite{Bethge2002,Berens2011,Mathis2012a}. In multi-scale codes,
the ratio of successive scales $\lambda_{k+1}/\lambda_k$
determines whether the Fisher information is appropriate:  if this ratio satisfies
Eq.~\ref{eq:nesting} with $S > 5$, the Fisher information estimate will be tight.
Here we showed how maximum likelihood estimation for a multi-scale population code
with circular Gaussian (von Mises) tuning curves (Eq.~\eqref{tuningcurves})
can be solved exactly, providing independent confirmation of the ideal scaling
ratios.  For each response, 
one can thus  not only estimate the most likely stimulus, but also gauge how reliable
the estimate is---the Fisher information only measures
the average reliability under ideal conditions.  If the stimulus comes from the
senses, such as vision or hearing, 
 the range of stimulus intensities can cover many orders of magnitude;  by having
neurons in a
population with different scales of sensitivity, the network can match its response
 to the dynamic range of the stimulus.   But even when the dynamic stimulus range 
does not extend over many orders of magnitude, the explicit maximum likelihood
solution shows that a multi-scale code is feasible, straightforwardly decodable, and
advantageous. 

Furthermore, we demonstrated that noise correlations at each scale reduce
the Fisher information in a nested grid code, but that the resolution still scales as
$J^L$, where $J$ is
the Fisher information of a single module and $L$ is the number of modules (each with
a different scale). Within a single module, the resolution scales linearly with the
number of neurons $M$, with or without noise
correlations~\cite{Shamir2006,Ecker2011}.


We measured the noise correlations in the spike trains from grid cells in the 
medial entorhinal cortex (mEC) of rats running on  a one-dimensional, linear
track~\cite{Hafting2008}.
 As in other studies in the visual~\cite{Ecker2010} and olfactory
cortices~\cite{Miura2012} we find that the
mean noise correlation of a random pair of neurons practically vanishes. However,
when two grid cells have highly
overlapping firing fields, so that both the spatial period and the phase are similar,
 the noise
correlations reach values of up to $0.8$.  We can imagine four different causes
for such comparatively high correlations: (i) grid cells in mEC mutually entrain to
the 
theta rhythm, a 5-12 Hz network rhythm present throughout hippocampus, subiculum, and
the entorhinal cortex.
 Grid cell spikes precess with respect to this rhythm, so that the spike phase
relative to the theta rhythm  shifts continuously and predictably, from the time that
the animal enters a firing field of a cell 
 until the time  it leaves that field~\cite{Hafting2008,Reifenstein2012}. If two
neurons have overlapping firing fields, similar theta phase precession could
 lead to high noise correlation. 
(ii) common external input (e.g. from hippocampus~\cite{Bonnevie2013},
or from other brain areas~\cite{VanStrien2009}) (iii) recurrent intrinsic
connections
between neurons in mEC (e.g.~\cite{Beed2010,Burgalossi2011,Couey2013,Beed2013}) (iv)
the spike trains have been acquired by spike-sorting extracellular
recordings~\cite{Hafting2008}; this process possibly falsely assign spikes from the
same neuron to different neurons and vice versa~\cite{Lewicki1998}, which can lead to
spurious
noise correlations.

Noise correlations are higher between neurons with similar tuning curves; this  can
have a strong effect on the coding precision of population
codes -- and nested grid codes are no exception.  Previous
studies focused on ensembles of cells with unimodal tuning
curves~\cite{Abbott1999,Shamir2001,Wilke2001,Shamir2006,Ecker2011}
rather than ensembles with multimodal tuning curves, such as grid cells.  
Interestingly, if one makes unimodal tuning curves heterogeneous by varying the
tuning widths and peak firing rates across neurons,
reducing the noise correlations does not improve encoding
accuracy~\cite{Wilke2001,Shamir2006,Ecker2011}. Grid cells are
also highly heterogeneous in their firing rates and tuning
parameters~\cite{Hafting2005,Herz2011}, but how heterogeneity would affect the Fisher
information has not been studied.

We have assumed that the  signal  itself is not subject to noise. If one adds
adiabatic noise to $x$, so that  $y=x+\xi$, 
then the population  encodes $y$ instead of $x$ to a certain precision, and 
one obtains 
$
  \left\langle { \left( x-\hat{x}\right)}^2\right\rangle \sim J^{-1}  + \left\langle
\xi^2 \right\rangle
$.  
While a coding strategy that uses multiple scales will still be superior, the law of
diminishing returns applies---reducing $J^{-1}$ far below $ \left\langle \xi^2
\right\rangle$
makes little sense. It is intuitive that the resolution at the coarsest scale should
limit the next length scale, but the resolution itself is sensitive to many
parameters,
such as  the peak firing rate and the number of neurons with tuning curves at the
coarsest scale. To minimize the number of spurious solutions in decoding the neuronal
response, a conservative approach might involve taking $1/2 < \lambda_{k+1}/\lambda_k
< 1$, independently of the resolution at scale $\lambda_k$. Maximum likelihood
decoding of a multi-scale population code relies on the constructive interference of
spatial oscillations  with periods  $\lambda_k$ and $\lambda_{k+1}$, which combine
near the
true stimulus $x$ to yield a high posterior
probability of the stimulus given the response (see 
Eq.~\eqref{eq:multi_scale_posterior}); any other instances of constructive
interference can lead to decoding errors. If the Fisher information is at least 
 $\sqrt{J} \sim \lambda_k$, then limiting  $1/2 < \lambda_{k+1}/\lambda_k < 1$ will
limit the probability of the  oscillations on the two scales constructively
interfering again  
 within $\lambda_k$ of the true stimulus $x$.   The ratio of scales observed in
experimental data seems to be in accordance with $0.6 < \lambda_{k+1}/\lambda_k
<0.8$~\cite{Barry2007,Stensola2012}.

Other authors have suggested that the different periods should not be simple
multiples of each
other~\cite{Burak2006,Gorchetchnikov2007,Fiete2008,Sreenivasan2011}; 
in the absence of noise,  each $x$ would then give rise to a unique pattern of 
population activity $\Upsilon$, 
up until $x$ reaches the least common multiple of all  the periods.  Hence, a much
larger range of $x$ can 
be encoded. Such a strategy is called modular arithmetic~\cite{Burak2006}; its main
drawback, though, is its susceptibility to
noise~\cite{Fiete2008,Sreenivasan2011,Mathis2012a}.
When $x$ evolves continuously in time, error correction could be used to counteract
this noise~\cite{Sreenivasan2011}. But even if $x$ has no continuous history 
dependence, a modular arithmetic code is feasible, provided $M$ is sufficiently
large---the specific model of Eq.~\eqref{tuningcurves} is explicitly solved by
Eq.~\eqref{eq:multi_scale_posterior}
 for any
set of spatial
periods $\lambda_i$, so one can make the expected error arbitrarily small as long as
one increases $N$. The authors of  Ref.~\cite{Wei2013} set themselves the opposite
goal
and try to minimize $N$; by using scaling arguments and dimensional analysis, they
derive optimal parameters for a multi-scale code.

These authors and we have treated the comparatively straightforward  problem of
encoding $x \in I  \subset \mathds{R}^D$ using multiple scales.
If instead the neuronal population represents the probability of a
stimulus $p(x)$ instead of just  the estimate of $x$~\cite{Pouget2000},  
then a
multi-scale encoding 
becomes analogous  to a Fourier decomposition of the probability distribution, given
the similarity 
between the set of periodic tuning curves at different length scales and the Fourier
basis. The analogy is approximate, though, as the 
corresponding "Fourier coefficients" would be highly stochastic given the inherent
randomness in the neuronal response;
moreover,  the set  of tuning curves is not complete.  A more detailed analysis of
probabilistic coding models  
in the context of multiple scales awaits investigation; our result that the
uncertainty in the position estimate 
fluctuates strongly as a function of the response, may be a first step in this
direction.

\section*{Acknowledgments}

We are thankful to the Moser lab from the Norwegian University of Science and
Technology for granting us access to the
grid cell data
and Christian Leibold for helpful discussions. This work was supported by the Federal
Ministry for Education and
Research (through the Bernstein Center for Computational Neuroscience Munich).

\end{document}